\documentclass[lettersize,journal]{IEEEtran}
\usepackage{amsmath,amsfonts}
\usepackage{algpseudocode}
\usepackage{algorithm}
\usepackage{array}
\usepackage{textcomp}
\usepackage[caption=false,font=normalsize,labelfont=sf,textfont=sf]{subfig}
\usepackage{tabularx}
\usepackage{stfloats}
\usepackage{url}
\usepackage{verbatim}
\usepackage{booktabs}
\usepackage{graphicx}
\usepackage{subcaption}
\usepackage{cite}
\usepackage{float}
\usepackage{pdfpages}
\usepackage{svg}
\usepackage{mathrsfs}
\usepackage{amssymb}
\usepackage{multirow}

\begin{document}
\title{Collaborative Computation and Migration in Multi-UAV Networks with Semantic Communication}

\author{Bin Li, Yuchen Ou, Yinqiu Liu, and Abbas Jamalipour,~\IEEEmembership{Fellow,~IEEE}

\thanks{Bin Li and Yuchen Ou are with the School of Computer Science, Nanjing University of Information Science and Technology, Nanjing 210044, China (e-mail: bin.li@nuist.edu.cn; 202412492169@nuist.edu.cn).}
\thanks{Yinqiu Liu is with the College of Computing and Data Science, Nanyang Technological University, Singapore 639798 (e-mail: yinqiu001@e.ntu.edu.sg).}
\thanks{Abbas Jamalipour is with School of Electrical and Computer Engineering, The University of Sydney, Sydney, Australia (e-mail: a.jamalipour@ieee.org).}
}


\maketitle
\begin{abstract}
Multi-Uncrewed Aerial Vehicle (UAV)-assisted Mobile Edge Computing (MEC) is a key technology for future 6G networks, providing wide coverage and flexible computing services. However, the limited resources of UAVs and the dynamic changes in the network structure make it difficult to maintain high efficiency. Existing methods often ignore the semantic information of tasks and the complex relationships among UAVs and mobile terminals, resulting in poor coordination. This paper proposes a joint optimization framework for task offloading, task migration, and trajectory control in semantic communication-enabled multi-UAV edge computing systems, aiming to maximize semantic similarity while minimizing task latency and system energy consumption. To tackle the resultant problem, we develop a Heterogeneous Graph Attention Network-based Multi-Agent Twin Delayed Deep Deterministic Policy Gradient (HAN-MATD3) algorithm. Specifically, we use a heterogeneous graph to model the network topology and apply HAN to extract important semantic features. These features are then integrated into the MATD3 framework to support joint optimization. Simulation results show that HAN-MATD3 improves the converged reward by 5.3\% over MATD3 (and 17.7\% over MADDPG), and reduces average latency by up to 12.2\%–14.6\% under the bandwidth and CPU settings, while maintaining higher semantic similarity.
\end{abstract}

\begin{IEEEkeywords}
Mobile edge computing, uncrewed aerial vehicle, task offloading, task migration, heterogeneous graph attention network, multi-agent deep reinforcement learning.
\end{IEEEkeywords}

\section{Introduction}

With the explosive growth of Internet of Things devices and the emergence of latency-sensitive applications such as autonomous driving and extended reality, traditional terrestrial networks face severe pressure in providing low-latency services \cite{MEC_survey, Guo2024MultiUAV}. Uncrewed Aerial Vehicles (UAVs) are flexible and can establish clear Line-of-Sight (LoS) communication links with ground Mobile Terminals (MTs). UAV-assisted Mobile Edge Computing (MEC) allows for enhancing network coverage and providing computing capabilities \cite{Hao2024MutliUAV}. However, the traditional Shannon theory-based communication paradigm, which focuses on the accurate transmission of bit sequences, is approaching its physical limits and struggles to support the massive data generated by intelligent tasks \cite{Shannon_limit}. Consequently, transmitting large volumes of raw data consumes too much energy and causes high latency. Therefore, in such multi-UAV MEC systems, a key challenge is to jointly optimize task latency and energy consumption while ensuring reliable data transmission under dynamic topology and limited onboard resources.

To address these challenges, semantic communication has emerged as a promising solution \cite{Hu2025SemUAV, Yan2024QoE}. Unlike traditional communication that transmits all bits, semantic communication extracts and sends only the semantic information of the data. This method can significantly reduce the data size without losing essential information. By applying semantic communication to UAV-assisted MEC systems, we can save transmission resources and reduce latency \cite{Zheng2025SemUAV, Yan2026SemanticMEC}. This makes semantic communication particularly suitable for resource-constrained UAV networks. To fully exploit these advantages, recent studies have begun to integrate semantic transmission models into wireless systems. As demonstrated by \cite{Shannon_limit}, semantic communication models can reliably transmit core meanings with minimal spectrum consumption. Furthermore, the work in \cite{Yan2024QoE} proposed semantic-aware resource allocation schemes to maximize the quality of experience in multi-task networks. These preliminary efforts validate that shifting from bit-level to semantic-level transmission is a viable strategy for alleviating the energy and latency constraints in modern edge computing scenarios.

Although semantic communication is effective, applying it to multi-UAV systems entails significant challenges. First, the high mobility of UAVs and MTs necessitates dynamic task migration to ensure service continuity, adding a layer of complexity to the joint optimization of the UAV trajectory and resource allocation \cite{Zhao2025Mobility, Yan2026UAVDelivery, Jia2026UAVSwarms}. Second, unlike bit-level transmission, the quality of semantic communication is highly sensitive to channel conditions. This makes topology-aware decision-making essential for maintaining high semantic similarity \cite{Hu2025SemUAV}. Existing solutions struggle to simultaneously capture the dynamic network topology and heterogeneous node interactions. Traditional Deep Reinforcement Learning (DRL) methods typically treat the network state as unstructured feature vectors, failing to capture the complex topological dependencies required for topology-aware scheduling \cite{Bian2023GNN}. While some graph-based reinforcement learning methods have been proposed \cite{Li2023GNNDRLUAV, Wang2024GNNDRL}, they often assume homogeneous nodes, thereby neglecting the distinct attributes and roles of UAVs and MTs. Consequently, these methods fail to effectively model the heterogeneous interactions in the network, leading to inefficient collaboration and sub-optimal performance.

To address these issues, we design a topology-aware cooperative decision-making framework for semantic-enabled multi-UAV MEC systems. The key idea is to represent the dynamic UAV-MT network as a heterogeneous graph instead of an unstructured state vector. In this graph, UAVs and MTs are modeled as different types of nodes, which enables the framework to capture spatial dependencies and heterogeneous interactions. We then integrate Heterogeneous Graph Attention Network (HAN) \cite{Wang2019HAN,Li2026Hierarchical} with Multi-Agent Twin Delayed Deep Deterministic Policy Gradient (MATD3) \cite{Zhao2022MATD3} under the CTDE paradigm. The HAN encoder extracts topology-aware node embeddings, and MATD3 uses these embeddings to jointly optimize UAV trajectory, task offloading, task migration, and resource allocation. To evaluate semantic transmission quality, we adopt an empirical semantic similarity model that reflects the impact of channel conditions and transmission parameters. The objective is to maximize semantic similarity while minimizing task execution latency and system energy consumption. The main contributions of this paper are summarized as follows:

\begin{itemize}
	\item Unlike existing studies focusing on bit-level transmission, we formulate a unified semantic communication-enabled multi-UAV MEC framework. The UAVs provide edge computing services to MTs while establishing Air-to-Ground (A2G) communication links for semantic information transmission. In this scenario, UAVs must coordinate task processing and dynamically adjust their trajectories to ensure favorable channel conditions, which leads to an intricate optimization challenge that simultaneously addresses task offloading, service migration, UAV trajectory design, and the distribution of resources.
	\item Considering the heterogeneous nature of UAVs and MTs alongside dynamic network topology, we propose a heterogeneous graph-based approach for better state representation. Specifically, we model the problem as a Partially Observable Markov Decision Process (POMDP) and employ a heterogeneous graph to capture distinct node features and varied cooperative relationships. This enables topology-aware decision-making that explicitly accounts for spatial relationships and semantic transmission quality.
	\item To tackle the resulting non-convex mixed-integer nonlinear programming problem, we propose an efficient HAN-MATD3 algorithm. Our algorithm integrates a HAN to extract features from the heterogeneous graph structure, and adopts the MATD3 to learn cooperative policies under the Centralized Training with Decentralized Execution (CTDE) paradigm. Compared with conventional vector-based DRL methods, HAN-MATD3 more effectively addresses the challenges of multiple constraints, large-scale variables, and dynamic environments in our multi-UAV semantic MEC scenario.
\end{itemize}

To further clarify the novelty of this work, Table~\ref{tab:related_comparison} compares the proposed framework with representative related studies. Existing UAV-MEC studies mainly focus on task offloading, resource allocation, or trajectory design under bit-level communication models. Some recent works have investigated semantic communication or graph-based DRL, but they usually did not jointly consider semantic transmission, task migration, UAV trajectory control, heterogeneous graph modeling, and multi-agent DRL. In contrast, our work integrates semantic communication with a heterogeneous graph-based MATD3 framework to jointly optimize task offloading, task migration, trajectory control, and resource allocation decisions.

\begin{table*}[t]
	\centering
	\caption{Comparison with representative related studies.}
	\label{tab:related_comparison}
	\begin{tabular}{lccccccc}
		\toprule
		Work & UAV-MEC & Semantic Communication & Offloading & Migration & Trajectory & HAN & MADRL \\
		\midrule
		Hao \textit{et al}. \cite{Hao2024MutliUAV} & Yes & No & Yes & No & Yes & No & No \\
		Pervez \textit{et al}. \cite{Pervez2024MutliUAV} & Yes & No & Yes & No & Yes & No & No \\
		Li \textit{et al}. \cite{Li2025DistributedMEC} & Yes & No & Yes & No & Yes & No & Yes \\
		Hu \textit{et al}. \cite{Hu2025SemUAV} & Partial & Yes & No & No & Yes & No & Partial \\
		Wang \textit{et al}. \cite{Wang2024GNNDRL} & Partial & No & Yes & Partial & No & No & No \\
		Proposed & Yes & Yes & Yes & Yes & Yes & Yes & Yes \\
		\bottomrule
	\end{tabular}
\end{table*}

The rest of this article is structured as follows: Section II provides a review of pertinent literature. Section III details the system architecture and defines the problem. The proposed HAN-MATD3 framework, encompassing the POMDP model and heterogeneous graph state representation, is presented in Section IV. Section V discusses the experimental configuration and evaluation outcomes. The paper is concluded in Section VI.

\section{Related Work}

\subsection{Multi-UAV-Assisted MEC}
Multi-UAV-assisted MEC is widely adopted as a flexible way to provide computing services and improve network coverage for ground users. Hao \textit{et al}. \cite{Hao2024MutliUAV} studied task offloading in multi-UAV MEC systems and improved convergence performance and latency. Pervez \textit{et al}. \cite{Pervez2024MutliUAV} jointly optimized task offloading, UAV paths, and resource use to reduce energy consumption and latency. Liu \textit{et al}. \cite{Liu2025UAV} developed a DRL-based framework for 3D UAV path planning and task offloading. Recent low-altitude wireless networking studies further extend this line of research. Jia \textit{et al}. \cite{Jia2026LowAltitudeATM} investigated hierarchical low-altitude network management for air traffic control, while Jia \textit{et al}. \cite{Jia2026RobustOffloading} studied distributionally robust computation offloading and trajectory optimization under low-altitude network uncertainty. Distributed many-agent optimization has also attracted increasing attention in UAV-driven MEC systems. Li \textit{et al}. \cite{Li2025DistributedMEC} jointly optimized UAV trajectories, UAV-MT associations, time-slot slicing, MT offloading powers, and MT local CPU clock speeds with the objective of maximizing expected energy efficiency, and analyzed the complexity, communication overhead, and convergence performance. However, most existing studies still rely on bit-level communication models and do not explicitly model semantic transmission or heterogeneous UAV-MT interactions. Different from these low-altitude networking and distributed MEC studies, this paper integrates semantic communication with heterogeneous graph representation and MATD3 to jointly optimize task offloading, task migration, trajectory control, and resource allocation.
\subsection{Task Offloading and Migration}

Due to the high mobility of UAVs and users, task migration is important for ensuring service continuity and load balancing in MEC networks \cite{Liu2022AoI}. Specifically, Wang \textit{et al}. \cite{Wang2023MigMEC} developed a user-centric service migration framework based on POMDP, using a long short-term memory-based encoder and an off-policy actor-critic algorithm to enable online decision-making with incomplete information. Furthermore, Peng \textit{et al}. \cite{Peng2024MigVehMEC} proposed a service migration scheme considering both computing and communication costs, using fast transfer reinforcement learning to maximize satisfaction while minimizing latency in dynamic vehicular edge computing networks. Nevertheless, traditional migration schemes typically involve transmitting the entire raw data or task context. Given the limited bandwidth of UAVs, this causes excessive communication overhead.

\subsection{Semantic Communication}

Semantic communication has risen as a promising solution to circumvent data transmission constraints by exclusively extracting and conveying task-pertinent information. For example, Hu \textit{et al}. \cite{Hu2025SemUAV} proposed a multi-UAV hybrid decision-controlled DRL scheme for multi-modal semantic communication networks, which jointly optimizes UAV trajectory and resource allocation to maximize semantic-aware Quality of Experience (QoE) while minimizing transmission costs. By utilizing a DRL approach, Wang \textit{et al}. \cite{Wang2023SemUAV} explored the joint design of resource management and UAV flight paths to boost the efficiency of semantic transmission. Furthermore, Zheng \textit{et al}. \cite{Zheng2025SemUAV} proposed a resource scheduling approach based on PER-SD3 for UAV-assisted MEC systems using semantic communication, which jointly optimizes UAV trajectory and task offloading to maximize QoE against malicious jamming attacks. However, these works either focus on point-to-point scenarios or single-UAV deployments with dedicated base stations, lacking consideration of multi-UAV collaborative edge computing where task migration and heterogeneous node interactions are essential.

\subsection{Graph Neural Network and Deep Reinforcement Learning}

Multi-Agent Deep Reinforcement Learning (MADRL) combined with Graph Neural Networks (GNN) has shown great potential in solving coordination problems in dynamic networks \cite{Multi2021Wang}. Some pioneering works have applied graph learning to wireless networks \cite{HRL2023Pamuklu,Zhang2023Cooperative}. Wang \textit{et al}. \cite{Wang2024GNNDRL} integrated GNNs with DRL algorithms to address task offloading and service management in MEC systems, effectively using topological dependencies to minimize system costs. Wu \textit{et al}. \cite{Wu2021GNNDRL} proposed a fine-grained task scheduling scheme combining GNN and DRL to optimize computation offloading for diverse task dependency topologies, aiming to minimize application completion time and energy consumption. However, conventional MADRL approaches typically treat the network state as unstructured feature vectors. They fail to effectively capture the heterogeneous interactions among different types of nodes, leading to sub-optimal collaboration policies.

\subsection{Contributions of This Paper}

To fill these research gaps, this paper proposes a joint computation offloading and task migration framework for multi-UAV-assisted MEC systems. Unlike existing works, we integrate semantic communication to reduce the overhead of task offloading and migration. Moreover, we design an HAN-MATD3 algorithm. By explicitly modeling the complex relationships among heterogeneous nodes, our approach jointly optimizes UAV trajectories, task offloading, task migration, and resource allocation strategies to minimize task latency and system energy consumption while maximizing semantic similarity.
\begin{table}[t]
	\caption{Summary of Key Notations}
	\label{tab:notations}
	\centering
	\begin{tabularx}{\linewidth}{lX}  
		\toprule
		\textbf{Notation} & \textbf{Description} \\
		\midrule
		$T$ & Total number of time slots \\
		$\Delta t$ & Duration of time slot \\
		$N,M$ & Number of UAVs and MTs\\
		$\mathcal{N}_n(t)$ & Sets of UAVs within the coverage of UAV $n$ \\
		$\mathcal{M}_n(t)$ & Sets of MTs within the coverage of UAV $n$ \\
		$d_{n,n'}(t)$ & Distance between UAVs \\
		$d_{n,m}(t)$ & Distance between UAV and MT \\
		$\mathbf{q}_n(t),\mathbf{q}_m(t)$ & Position vector of UAV $n$ and MT $m$ \\
		$I_{n,m}(t)$ & Binary task offloading decision variable \\
		$J_{n,n',m}(t)$ & Binary task migration decision variable \\
		$v_n(t), \theta_n(t)$ & Flight speed and direction vector of UAV $n$ \\
		$W_{n,m}(t)$ & Bandwidth allocated for task offloading \\
		$W_{n,n'}(t)$ & Bandwidth allocated for task migration \\
		$f_{n,m}(t)$ & Computation resource allocation \\
		$C^{\mathrm{com}}$ & Computation cycles per bit for task processing \\
		$C^{\mathrm{enc}}$ & Computation cycles per bit for semantic encoding \\
		$C^{\mathrm{dec}}$ & Computation cycles per bit for semantic decoding \\
		$u_m(t)$ & Data size of the task generated by MT $m$ \\
		$r_{n,m}(t), r_{n,n'}(t)$ & SINR for A2G and A2A links \\
		$n_b$ & Number of quantization bits \\
		$\delta(t)$ & Average semantic similarity in time slot $t$ \\
		$L^{\mathrm{total}}(t)$ & Total system latency in time slot $t$ \\
		$E^{\mathrm{total}}(t)$ & Total system energy consumption in time slot $t$ \\
		\bottomrule
	\end{tabularx}
\end{table}
\section{System Model}

This section first describes the system model and subsequently formulates the optimization problem. Table~\ref{tab:notations} provides a summary of the key notations utilized throughout this paper.

\subsection{Network Architecture}

\begin{figure}[t]
	\centering
	\includegraphics[width=\columnwidth]{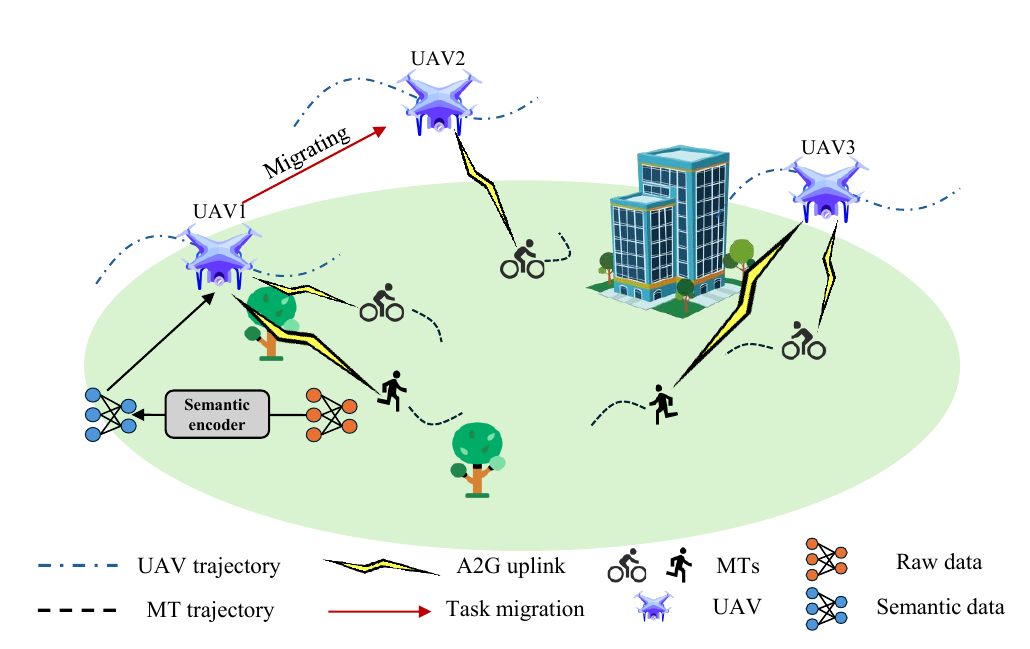}
	\caption{Illustration of the semantic communication-enabled multi-UAV MEC system.}
	\label{fig:system_model}
\end{figure}

As illustrated in Fig. \ref{fig:system_model}, this paper investigates a multi-UAV-assisted MEC system enabled by semantic communication, comprising UAVs and MTs. We consider a set of $N$ UAVs, denoted by $\mathcal{N}=\{1,2,\ldots,N\}$, flying within a constrained altitude range. There are $M$ MTs denoted by the set $\mathcal{M}=\{1,2,\ldots,M\}$, which move randomly on the ground. The service period is partitioned into $T$ equal time slots indexed by $t\in\{1,2,...,T\}$. In each time slot $t$, each MT $m$ generates a computation task with a data size of $u_m(t)$. 

We define a binary task offloading decision variable $ I_{n,m}(t)\in\{0,1\}$. Specifically, $ I_{n,m}(t)=1$ indicates that the task of MT $m$ is offloaded to UAV $n$ at time slot $t$; otherwise, the task is executed locally. Furthermore, we define a binary migration decision variable $J_{n,n',m}(t)\in\{0,1\}$. Here, $J_{n,n',m}(t)=1$  signifies that the task of MT $m$ is migrated from UAV $n$ to UAV $n'$ at time slot $t$.

Similarly, we define $\Omega_{n,m}^{\mathrm{Cov}}(t)\in\{0,1\}$ and $\Omega^{\mathrm{Cov}}_{n,n'}(t)\in\{0,1\}$ to represent whether MT $m$ or UAV $n'$ is within the coverage area of UAV $n$ in time slot $t$, respectively.  The sets $\mathcal{M}_n(t)$ and $\mathcal{N}_n(t)$ denote the MTs and the UAVs that fall within the coverage area of UAV $n$ in time slot $t$.

\subsection{Mobility Model}

We consider a 3D coordinate system. The position of UAV $n$ in time slot $t$ is denoted by $\mathbf{q}_n(t)=[x_n(t),y_n(t),z_n(t)]^T$, where $x_n(t)$, $y_n(t)$, $z_n(t)$ are the X, Y, and Z coordinates of UAV $n$, respectively. We denote $\mathbf{\theta}_n^\mathrm{UAV}(t)=[\theta_n^\mathrm{H}(t),\theta_n^\mathrm{V}(t)]$ as the flight direction of UAV $n$ in time slot $t$, where $\theta_n^\mathrm{H}(t)$ and $\theta_n^\mathrm{V}(t)$ represent the horizontal and vertical flight angles, respectively. Let $v_n(t)$ denote the flight speed of UAV $n$. Consequently, the position of UAV $n$ at the next time slot $\mathbf{q}_n(t+1)$ can be updated as
\begin{equation}
    \mathbf{q}_n(t+1)=\begin{bmatrix}x_n(t)+v_n(t)\cos(\theta_n^\mathrm{V}(t))\cos(\theta_n^\mathrm{H}(t))\\y_n(t)+v_n(t)\cos(\theta_n^\mathrm{V}(t))\sin(\theta_n^\mathrm{H}(t))\\z_n(t)+v_n(t)\sin(\theta_n^\mathrm{V}(t))\end{bmatrix}.
\end{equation}

The distance between UAVs $n$ and UAV $n'$ is given by $d_{n,n'}(t)=\|\mathbf{q}_n(t)-\mathbf{q}_{n'}(t)\|$. To ensure flight safety and operational constraints, the following conditions are subject to
\begin{equation}
    d_{n,n'}(t) \geq d_\mathrm{min},\label{safely_cs}
\end{equation}
\begin{equation}
    z_{\mathrm{min}}\leq z_n(t)\leq z_{\mathrm{max}},\label{altitude_cs}
\end{equation}
where $d_\mathrm{min}$, $z_\mathrm{min}$ and $z_\mathrm{max}$ denote the minimum safety distance, minimum altitude, and maximum altitude, respectively.

For MTs, we adopt the Gauss-Markov mobility model \cite{Liu2020GM}. The position of MT $m$ in time slot $t$ is denoted by $\mathbf{q}_m(t)=[x_m(t),y_m(t),0]^T$, where $x_m(t)$, $y_m(t)$ are the X and Y coordinates of MT $m$, respectively. Given its speed $v_m(t)$ and movement direction $\theta_m(t)$, the next position $\mathbf{q}_m(t+1)$ is expressed as
\begin{equation}
\mathbf{q}_m(t+1)=\begin{bmatrix}x_m(t)+v_m(t)\cos(\theta_m(t))\\y_m(t)+v_m(t)\sin(\theta_m(t))\\0\end{bmatrix}.
\end{equation}

Finally, the positions of all UAVs and MTs are required to remain within the simulation area, which can be formulated as
\begin{equation}
    0\leq x_n(t),y_n(t)\leq L,\label{area_cs_1}
\end{equation}
\begin{equation}
    0\leq x_m(t),y_m(t)\leq L,\label{area_cs_2}
\end{equation}
where $L$ is the side length of the simulation area.

\subsection{Channel Model}

We consider two types of communication links: A2G for task offloading and Air-to-Air (A2A) for task migration. Each has its own bandwidth pool.

\subsubsection{A2G channel model}
The A2G channel follows the commonly used probabilistic LoS A2G model for urban UAV communications \cite{AlHourani2014LAP}. The distance between UAV $n$ and MT $m$ in time slot $t$ is $d_{n,m}(t)=\|\mathbf{q}_n(t)-\mathbf{q}_m(t)\|$. In time slot $t$, the probability of the LoS link from UAV $n$ to MT $m$ can be expressed as
\begin{equation}
    \mathrm{P}_{\mathrm{LoS}}(t)=\frac1{1+\kappa_0\exp[-\kappa_1(\theta_{n,m}(t)-\kappa_0)]},
\end{equation}
where $\kappa_0$ and $\kappa_1$ are constants related to the environment, and $\theta_{n,m}(t)$ denotes the elevation angle from UAV $n$ to MT $m$, which can be expressed as
\begin{equation}
    \theta_{n,m}(t)=\frac{180}{\pi}\arcsin\left(\frac{z_n(t)}{d_{n,m}(t)}\right).
\end{equation}
The probability of the Non-Line-of-Sight (NLoS) link from UAV $n$ to MT $m$ is $\mathrm{P}_{\mathrm{NLoS}}(t)=1-\mathrm{P}_{\mathrm{LoS}}(t)$. The path losses of the LoS link and the NLoS link are
\begin{equation}
    \mathrm{L}_{\mathrm{LoS}}(t)=\left(\frac{4\pi f_\mathrm{c}d_{n,m}(t)}c\right)^\lambda\eta_\mathrm{LoS},
\end{equation}
\begin{equation}
    \mathrm{L}_{\mathrm{NLoS}}(t)=\left(\frac{4\pi f_\mathrm{c}d_{n,m}(t)}c\right)^\lambda\eta_\mathrm{NLoS},
\end{equation}
where $f_c$ denotes the center carrier frequency, $\lambda$ is the path loss exponent, $c$ is the speed of light, $\eta_\mathrm{LoS}$ and $\eta_\mathrm{NLoS}$ denote the excessive path loss of the LoS link and the NLoS link,  respectively. The average channel gain is expressed as
\begin{equation}
    h_{n,m}(t)=\frac1{\mathrm{P}_{\mathrm{LoS}}(t)\mathrm{L}_{\mathrm{LoS}}(t)+\mathrm{P}_{\mathrm{NLoS}}(t)\mathrm{L}_{\mathrm{NLoS}}(t)}.
\end{equation}
Considering the interference from other MTs within the coverage area to the uplink, the Signal-to-Interference-plus-Noise Ratio (SINR) for the uplink can be expressed as
\begin{equation}
    r_{n,m}(t)=\frac{h_{n,m}(t)p_m(t)}{\sum_{k\in\mathcal{M}_n,k\ne m}h_{n,k}(t)p_k(t)+\sigma^{2}},
\end{equation}
where $\sigma^{2}$ is the noise power, and $p_m(t)$ is the transmit power of MT $m$.

\subsubsection{A2A channel model}
Since UAVs operate at high altitudes, the A2A links are dominated by LoS propagation. In time slot $t$, the A2A channel gain from UAV $n$ to UAV $n'$ can be expressed as
\begin{equation}
    h_{n,n'}(t)=\frac{1}{\mathrm{L}_\mathrm{LoS}(t)}.
\end{equation}
Considering the interference from other UAVs within the coverage area to the link, the SINR between UAV $n$ and UAV $n'$ in time slot $t$ can be expressed as
\begin{equation}
    r_{n,n'}(t)=\frac{h_{n,n'}(t)p_n(t)}{\sum_{k\in\mathcal{N}_n,k\ne n}h_{k,n'}(t)p_k(t)+\sigma^{2}},
\end{equation}
where $p_n(t)$ is the transmit power of UAV $n$.

\subsection{Semantic Communication Transmission Model}

In semantic communication, MTs transmit semantic feature vectors extracted by a deep semantic encoder rather than traditional raw bits. For a raw task $U_m(t)$, the encoder maps the input data into a low-dimensional semantic feature space. These continuous features are then quantized into discrete semantic bits, denoted as
\begin{equation}
	b_m(t)=Q(F_{\mathrm{en}}(U_m(t))),
\end{equation}
where $F_{\mathrm{en}}(\cdot)$ represents the semantic encoding network, and $Q(\cdot)$ is the quantization operator. The number of quantization bits $n_b$ determines both the precision of the preserved semantic information and the upper bound of semantic performance.

The quantized semantic features $b_m$ are transmitted via the uplink channel. The achievable transmission rate between MT $m$ and UAV $n$ is given by
\begin{equation}
	R_{n,m}(t)=W_{n,m}(t)\log_2(1+r_{n,m}(t)),
\end{equation}
where $W_{n,m}(t)$ is the bandwidth allocated by UAV $n$ to MT $m$. Assuming the total bandwidth of UAV $n$ in time slot $t$ is $W_n(t)$, the bandwidth allocation is subject to
\begin{equation}
	\sum_{m\in\mathcal{M}_n}W_{n,m}(t)\leq W_n(t).\label{bandwidth_off_cs}
\end{equation}
Similarly, if task migration is triggered, the A2A transmission rate between UAV $n$ and UAV $n'$ is
\begin{equation}
	R_{n,n'}(t)=W_{n,n'}(t)\log_2(1+r_{n,n'}(t)).
\end{equation}
Assuming the total migration bandwidth of UAV $n$ in time slot $t$ is $W_n^\mathrm{A}(t) $, the bandwidth allocation is subject to
\begin{equation}
	\sum_{n'\in\mathcal{N}_n} W_{n,n'}(t) \le W_n^{\mathrm{A}}(t).\label{bandwidth_mig_cs}
\end{equation}

Upon receiving the semantic features, UAV $n$ performs de-quantization and uses a semantic decoder to reconstruct the original content. The decoded task is represented as
\begin{equation}
	\hat{U}_m(t)=F_{\mathrm{de}}(Q^{-1}(b_m(t))),
\end{equation}
where the goal is to maximize the semantic consistency between the reconstructed data $\hat{U}_m(t)$ and the original source $U_m(t)$.

Since deep learning-based encoders/decoders are non-linear, it is challenging to characterize the end-to-end semantic performance using Shannon capacity. Thus, we use the well-validated ABG model \cite{Ma2025semantic} to capture the dependence of semantic similarity on channel SINR. The adopted ABG model provides a tractable system-level abstraction of semantic recovery quality, which allows the semantic transmission process to be incorporated into the joint optimization of UAV trajectory, task offloading, task migration, and resource allocation. This formulation serves as an empirical semantic-quality approximation rather than a universal semantic encoder model. The semantic similarity for a task transmitted from MT $m$ to UAV $n$ is defined as
\begin{equation}
	\delta_{n,m}(t)=g_1(n_{b})-\frac{g_2}{1+\left(g_3 r_{n,m}(t)\right)^{g_4}},
\end{equation}
where $g_1(n_{b})$ represents the upper bound of semantic performance determined by quantization bits, and $g_2$, $g_3$, $g_4$ are fitting parameters that capture the sensitivity of the semantic model to channel quality. Consequently, the average system semantic similarity in time slot $t$ is
\begin{equation}
	\delta(t)=\frac{1}{M}\sum_{m\in\mathcal{M}}\sum_{n\in\mathcal{N}} I_{n,m}(t)\delta_{n,m}(t).
\end{equation}

\subsection{Latency Model}

For local task execution, only the computation latency is taken into account, since the encoding process is unnecessary. The local execution latency is given by
\begin{equation}
	L_m^{\mathrm{loc}}(t)=\frac{u_m(t)C^\mathrm{com}}{f_m(t)},
\end{equation}
where $u_m(t)$ is the size of the raw data $U_m(t)$, $C^\mathrm{com}$ denotes the number of CPU cycles required per bit, and $f_m(t)$ is the local CPU frequency of MT $m$.

For the tasks offloaded to the edge UAV, the total latency comprises four components: local encoding, uplink transmission, edge decoding, and edge computing. First, the raw task is semantically encoded locally. The encoding latency is
\begin{equation}
	L^\mathrm{enc}_m(t)=\frac{u_m(t)C^\mathrm{enc}}{f_m(t)},
\end{equation}
where $C^\mathrm{enc}$ represents the computational cycles per bit for semantic encoding at MT $m$. Next, the encoded features are transmitted. The transmission latency is
\begin{equation}
	L_{n,m}^\mathrm{tra}(t)=\frac{\hat{b}_m(t)}{R_{n,m}(t)},
\end{equation}
where $\hat{b}_m(t)$ is the size of semantic feature $b_m(t)$. Upon reception, UAV $n$ decodes the semantic information. The decoding latency is
\begin{equation}
	L_{n,m}^\mathrm{dec}(t)=\frac{\hat{b}_m(t)C^\mathrm{dec}}{f_{n,m}(t)},
\end{equation}
where $C^{\mathrm{dec}}$ is the computational cycles per bit for decoding, and $f_{n,m}(t)$ is the computation resource allocated by UAV $n$ to MT $m$. Assuming the total computation resource of UAV $n$ in time slot $t$ is $f_n (t)$, the resource allocation is subject to
\begin{equation}
	\sum_{m\in\mathcal{M}}f_{n,m}(t)\leq f_n(t).\label{com_cs}
\end{equation}
Finally, the reconstructed task $\hat{u}_m(t)$ is processed by the UAV. The computation latency is
\begin{equation}
	L_{n,m}^{\mathrm{com}}(t)=\frac{\hat{u}_m(t)C^\mathrm{com}}{f_{n,m}(t)},
\end{equation}
where $\hat{u}_m(t)$ is the size of $\hat{U}_m(t)$. The total offloading latency for MT $m$ served by UAV $n$ is
\begin{equation}
	L_{n,m}^{\mathrm{off}}(t)=L^\mathrm{enc}_m(t)+L_{n,m}^{\mathrm{tra}}(t)+L^\mathrm{dec}_{n,m}(t)+L_{n,m}^{\mathrm{com}}(t).
\end{equation}

If task migration occurs, the migration latency is treated as an additional transmission latency, that is,
\begin{equation}
	L_{n,n',m}^{\mathrm{mig}}(t)=\frac{\hat{b}_m(t)}{R_{n,n'}(t)}.
\end{equation}

In summary, the total latency for MT $m$ in time slot $t$ is formulated as
\begin{equation}
	\begin{aligned}
		L^{\mathrm{total}}_m(t)&=\left(1-\sum_{n\in\mathcal{N}}I_{n,m}(t)\right)L_m^\mathrm{loc}(t)\\&+\sum_{n\in\mathcal{N}}I_{n,m}(t)L^\mathrm{off}_{n,m}(t)\\&+\sum_{n\in\mathcal{N}}\sum_{n^{\prime}\in\mathcal{N}_n}J_{n,n',m}(t)L_{n,n^{\prime},m}^{\mathrm{mig}}(t).
	\end{aligned}
\end{equation}

Thus, the total system latency in time slot $t$ is expressed as
\begin{equation}
	L^{\mathrm{total}}(t)=\sum_{m\in\mathcal{M}}L^{\mathrm{total}}_m(t).
\end{equation}

\subsection{Energy Model}

For the tasks executed locally, the computation energy consumption at the MT can be modeled as
\begin{equation}
	E_m^\mathrm{loc}(t)=\kappa[f_m(t)]^3 L_m^\mathrm{loc}(t),
\end{equation}
where $\kappa$ is the effective switched capacitance coefficient depending on the CPU architecture. If a task is offloaded to a UAV, it must first be encoded at the MT. The encoding energy consumption is expressed as
\begin{equation}
	E_m^\mathrm{enc}(t)=\kappa[f_m(t)]^3 L_m^\mathrm{enc}(t).
\end{equation}

Therefore, the aggregate energy consumption of MT $m$ in time slot $t$ is given by
\begin{equation}
	\begin{aligned}
		E_m^\mathrm{total}(t)&=\left(1-\sum_{n\in\mathcal{N}} I_{n,m}(t)\right)E_m^\mathrm{loc}(t)\\&+\sum_{n\in\mathcal{N}} I_{n,m}(t)E_m^\mathrm{enc}(t).
	\end{aligned}
\end{equation}

Following \cite{Fotouhi2017UAVEnergy}, the communication energy consumption of a UAV is negligible compared to its propulsion energy. Thus, we focus on the energy consumed by flying, decoding, and computing. Based on \cite{Zeng2019UAVEnergy}, the propulsion energy consumption of UAV $n$ in time slot $t$ is modeled as
\begin{equation}
	\begin{aligned}
		E_n^{\mathrm{fly}}(t)&=\Delta t[P_0\left(1+\frac{3v_n(t)^2}{U_\text{tip}^2}\right)\\&+P_1\left(\sqrt{1+\frac{v_n(t)^4}{4v_0^4}}-\frac{v_n(t)^2}{2v_0^2}\right)^{1/2}\\&+\frac{1}{2}d_0\rho sAv_n(t)^3],
	\end{aligned}
\end{equation}
where $P_0$ and $P_1$ denote the profile power and the induced power required for hovering status, respectively. Furthermore, the environmental and hardware constants include the air density $\rho$, the rotor disc area $A$, the rotor solidity $s$, and the fuselage drag ratio $d_0$. Finally, $U_\mathrm{tip}$ and $v_{0}$ correspond to the rotor blade's tip speed and the average induced velocity during hover.

The energy consumption for semantic decoding and task computing at UAV $n$ is given by
\begin{equation}
	E_{n}^{\mathrm{com}}(t)=\sum_m\kappa[f_{n,m}(t)]^{3}(L_{n,m}^{\mathrm{com}}(t)+L_{n,m}^{\mathrm{dec}}(t)).
\end{equation}

The total energy consumption of UAV $n$ in time slot $t$ is expressed as
\begin{equation}
	E_n^{\mathrm{total}}(t)=E_n^{\mathrm{fly}}(t)+E_{n}^{\mathrm{com}}(t).
\end{equation}

Finally, the total system energy consumption in time slot $t$ is
\begin{equation}
	E^\mathrm{total}(t)=\sum_{m\in\mathcal{M}} E_m^\mathrm{total}(t)+\sum_{n\in\mathcal{N}} E_n^\mathrm{total}(t).
\end{equation}

\subsection{Problem Formulation}

We consider a multi-UAV assisted MEC system incorporating semantic communication. We aim to maximize semantic similarity while minimizing task latency and UAV energy consumption by jointly optimizing UAV trajectory, task offloading and migration decisions, and computation and communication resource allocation. The cost function is formulated as
\begin{equation}
	\begin{aligned}
		C(t)=\omega_{L}\bar{L}(t)+\omega_{E}\bar{E}(t)-\omega_{\delta}\bar{\delta}(t),
	\end{aligned}
\end{equation}
where the normalized latency, energy consumption, and semantic similarity are defined as
\begin{equation}
	\begin{aligned}
		\bar{L}(t)&=\frac{L^\mathrm{total}(t)}{T\Delta t},&
		\bar{E}(t)&=\frac{E^{\mathrm{total}}(t)}{E^\mathrm{batt}},&
		\bar{\delta}(t)&=\delta(t),
	\end{aligned}
\end{equation}
where $\omega_L$, $\omega_E$ and $\omega_\delta$ denote the weighting factors. The optimization problem is formulated as
\begin{subequations}
	\begin{align}     
		{ \min_{\substack{\mathbf{q}, \boldsymbol{I}, \boldsymbol{J},\\ \boldsymbol{B}, \boldsymbol{f}}} }& \quad {\sum_{t=1}^{T}C(t)} \label{eq:a} \\     
		\text{s.t.} 
		& \quad   
		\eqref{safely_cs}, \eqref{altitude_cs}, \eqref{area_cs_1}, \eqref{area_cs_2}, \eqref{bandwidth_off_cs}, \eqref{bandwidth_mig_cs}, \eqref{com_cs}\\     
		& \quad 
		0\leq v_n(t)\leq v^\mathrm{max},\ \forall n\in \mathcal{N}, t \label{eq:cons:speed},\\
		& \quad 
		\sum_{n\in\mathcal{N}} I_{n,m}(t)\leq 1,\ \forall m\in \mathcal{M},t\label{eq:cons:off}, \\
		& \quad 
		\sum_{n'\in\mathcal{N}}J_{n,n',m}(t)\leq 1,\forall m\in\mathcal{M},t \label{eq:cons:mig}, \\     
		& \quad 
		J_{n,n',m}(t)\leq I_{n,m}(t),\forall n\in\mathcal{N},m\in\mathcal{M},t\label{eq:cons:mig_2}, \\
		& \quad 
		\delta(t)\geq \delta^{\mathrm{th}},\forall  	t\label{eq:cons:sem}, \\
		& \quad 
		\sum_{t=1}^TE_n^\mathrm{total}(t)\leq E^\mathrm{batt},\forall n\in\mathcal{N} \label{eq:cons:power},
	\end{align}
\end{subequations}
where $v^\mathrm{max}$, $\delta^{\mathrm{th}}$, and $E^\mathrm{batt}$ denote the maximum flight speed of the UAV, the minimum semantic similarity threshold, and the total battery capacity, respectively. Constraints \eqref{area_cs_1} and \eqref{area_cs_2} restrict the mobility range of UAVs and MTs.  Constraints \eqref{safely_cs}, \eqref{altitude_cs} and \eqref{eq:cons:speed} specify the UAV's trajectory limits, including collision avoidance, altitude, and speed. Constraints \eqref{eq:cons:off}-\eqref{eq:cons:mig_2} ensure the validity of task offloading and migration decisions. Constraints \eqref{bandwidth_mig_cs}, \eqref{bandwidth_off_cs} and \eqref{com_cs} represent the resource budgets for migration bandwidth, communication bandwidth, and computation capacity, respectively. Finally, constraints \eqref{eq:cons:sem} and \eqref{eq:cons:power} guarantee the semantic reliability requirements and the UAV's energy constraints.

\section{Proposed Solution}

In this section, we present the proposed framework. First, we formulate the problem as a POMDP. Then, we introduce the heterogeneous graph construction to model the network topology. Next, we describe the graph feature encoding process based on HAN. Finally, we propose the HAN-MATD3 algorithm and provide its complexity analysis. 

\subsection{POMDP Formulation}

Due to the limited sensing range of UAVs, each agent only possesses partial visibility of the global environment. Therefore, we formulate the joint optimization problem as a POMDP, defined by the tuple $(\mathcal{N}, \mathcal{S}, \mathcal{A}, \mathcal{P}, \mathcal{O}, \mathcal{R}, \gamma)$, where $\mathcal{N}=\{1,2,...,N\}$ denotes the set of UAV agents and $\gamma\in[0,1]$ is the discount factor. Constraint \eqref{area_cs_2} regarding MTs is inherently satisfied by the simulation environment settings. The detailed components are defined as follows.

\subsubsection{State $\mathcal{S}$}
The global state $s(t)$ characterizes the complete environmental information in time slot $t$, which is available only during the centralized training phase. It consists of the states of all UAVs and MTs, expressed as $s(t)=\{\mathbf{s}_{1}^{\mathrm{UAV}}(t),\ldots,\mathbf{s}_{N}^{\mathrm{UAV}}(t),\mathbf{s}_{1}^{\mathrm{MT}}(t),\ldots,\mathbf{s}_{M}^{\mathrm{MT}}(t)\}$. The UAV state includes position, velocity, flight direction, remaining energy, and local computing-resource information, while the MT state includes position, task size, and channel-related features.

\subsubsection{Observation $\mathcal{O}$}
Due to the limited communication range, each UAV $n$ obtains only a local observation $o_n(t)=\{o_n^{\mathrm{self}}(t),o_n^{\mathrm{other}}(t),o_n^{\mathrm{mt}}(t)\}$. Here, $o_n^{\mathrm{self}}(t)$ contains the local mobility, energy, and resource state of UAV $n$; $o_n^{\mathrm{other}}(t)$ contains the relative states of neighboring UAVs; and $o_n^{\mathrm{mt}}(t)$ contains the task and link-state information of MTs within the communication range. These components form the input features of the HAN encoder and the actor-critic networks.

\subsubsection{Action $\mathcal{A}$} 
The action space is defined as $a_n(t)=\{a_n^{\mathrm{fly}}(t),a_n^{\mathrm{off}}(t),a_n^{\mathrm{mig}}(t),a_n^{\mathrm{res}}(t)\}$, where $a_n^{\mathrm{fly}}(t)$ represents the flight decision, $a_n^{\mathrm{off}}(t)$ and $a_n^{\mathrm{mig}}(t)$ denote the offloading and migration decisions, respectively, and $a_n^{\mathrm{res}}(t)$ represents the resource allocation decision.

To satisfy constraint \eqref{eq:cons:speed}, we restrict the velocity $v_n(t)$ within valid ranges using the tanh activation function. To address the binary constraints \eqref{eq:cons:off}, \eqref{eq:cons:mig}, and \eqref{eq:cons:mig_2} for task offloading and migration decisions, the actor network outputs continuous probability vectors. We then employ the argmax operation during the execution phase to select a unique target UAV for each task, ensuring that each task is processed by at most one UAV. However, to resolve the non-differentiability of argmax during the centralized training phase, we utilize the gumbel-softmax relaxation technique. This approximation allows gradients to backpropagate effectively from the critic to the actor, thereby ensuring the stability of the convergence process.  Furthermore, to satisfy constraints \eqref{bandwidth_off_cs}, \eqref{bandwidth_mig_cs}, and \eqref{com_cs}, we utilize the softmax function to determine the resource allocation ratios.

\subsubsection{Transfer probability $\mathcal{P}$}
Let $\mathcal{P}$ denote the state transition probability function. $\mathcal{P}(s(t+1)|s(t),a(t))$ represents the probability of the system transitioning from state $s(t)$ to $s(t+1)$ after the agents execute the joint action $a(t)$.

\subsubsection{Reward $\mathcal{R}$}
The reward function is formulated to steer the agents towards optimizing the objective function. It is composed of a global utility term and penalty terms, formulated as
\begin{equation}
	r(t) = -C(t)-(P_{\text{coll}}(t)+P_{\text{batt}}(t)+P_{\text{sem}}(t)).
\end{equation}

Since the reward is the negative form of the weighted cost, the latency term encourages UAVs to select offloading, migration, trajectory, and resource-allocation decisions that reduce end-to-end task execution time. The energy term discourages excessive propulsion and computation energy consumption. The semantic-similarity term provides a positive incentive for maintaining high semantic recovery quality under time-varying wireless channels. The penalty terms further guide the agents to satisfy safety, battery, and semantic reliability constraints during training.
To explicitly quantify constraint violations, we utilize the indicator function $\mathbb{I}(\cdot)$, which equals 1 if the condition is true and 0 otherwise. The collision and boundary penalties $P_{\text{coll}}(t)$ are mathematically formulated as
\begin{equation}
	\begin{aligned}
	P_{\text{coll}}(t)&=\vartheta_1 \sum_{n \neq n'} \mathbb{I}(d_{n,n'}(t) < d_{\min}) \\&+ \vartheta_2 \sum_{n} \mathbb{I}(\mathbf{q}_n(t) \notin \mathcal{S}_{\mathrm{area}}),
	\end{aligned}
\end{equation}
where $\mathcal{S}_{\mathrm{area}}$ denotes the valid flight area. Furthermore, the penalties for battery exhaustion $P_{\text{batt}}(t)$ and semantic reliability violation $P_{\text{sem}}(t)$ are respectively given by
\begin{equation}
	P_{\text{batt}}(t) = \vartheta_3 \sum_{n} \mathbb{I}(E_n(t) < 0),
\end{equation}
\begin{equation}
	P_{\text{sem}}(t) = \vartheta_4 \mathbb{I}(\delta(t) < \delta^{\text{th}}),
\end{equation}
where $\vartheta_1$, $\vartheta_2$, $\vartheta_3$ and $\vartheta_4$ represent the corresponding penalty coefficients. To maximize the cumulative reward, the agents learn to avoid these penalties during training, thereby satisfying the constraints of the optimization problem.

\subsection{Heterogeneous Graph Construction}

Traditional reinforcement learning methods typically treat input observations as simple feature vectors, which fail to capture the complex topological dependencies among agents. To address this limitation, we model the system state at each time slot as a heterogeneous graph $G\triangleq(\mathcal{V},\mathcal{E})$. The node set $\mathcal{V}$ contains two types of entities: UAVs and MTs. UAV nodes act as agents responsible for executing actions and communicating with other nodes. MT nodes serve as service recipients, providing task demands and location information. The edge set $\mathcal{E}$ explicitly captures the topological connections between nodes, including UAV-UAV edges reflecting collaboration and collision avoidance relationships \cite{Jia2026LowAltitudeATM}, and MT-UAV edges reflecting service requests and A2G link quality. At each time slot, the edge set is constructed according to the communication and coverage relationships. A UAV-UAV edge $(n,n')$ exists if $n\neq n'$ and $d_{n,n'}(t)\le R_{\mathrm{U2U}}$, where $R_{\mathrm{U2U}}$ denotes the UAV-to-UAV communication range. A UAV-MT edge $(n,m)$ exists if $d_{n,m}(t)\le R_{\mathrm{U2M}}$, where $R_{\mathrm{U2M}}$ denotes the UAV-to-MT coverage radius. Therefore, the neighbor sets used by the UAV-UAV and MT-UAV meta-paths are $\mathcal{N}_n(t)=\{n'\in\mathcal{N}, n'\neq n \mid d_{n,n'}(t)\le R_{\mathrm{U2U}}\}$ and $\mathcal{M}_n(t)=\{m\in\mathcal{M}\mid d_{n,m}(t)\le R_{\mathrm{U2M}}\}$, respectively. The relative position, LoS probability, and channel-gain features on these edges are then used by the attention mechanism to learn the importance of different neighboring entities.

Based on the graph topology, we define two types of meta-paths to guide the information aggregation process. The meta-path $\Phi_{nn}$ (UAV-UAV) directs agents to focus on neighboring UAV states, physically capturing the cooperative interference and collision avoidance relationships. The meta-path $\Phi_{mn}$ (MT-UAV) directs agents to focus on user service demands, capturing the service demand distribution for efficient task offloading and migration decision-making.

\begin{figure}[t]
	\centering
	\includegraphics[width=\columnwidth]{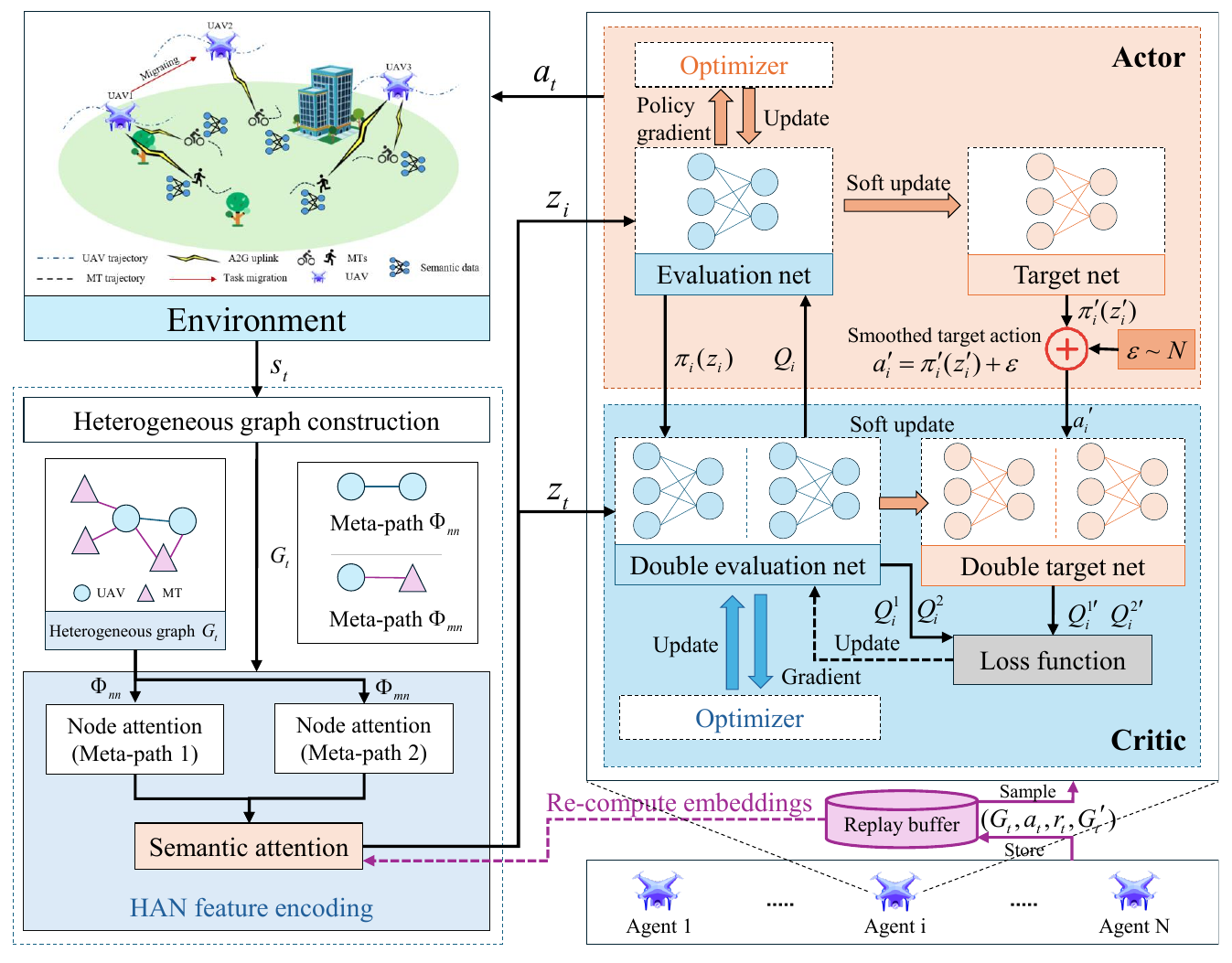}
	\caption{The overall framework of the proposed algorithm.}
	\label{fig:algorithm}
\end{figure}

\subsection{Graph Feature Encoding}

A heterogeneous graph encoder is adopted because UAVs and MTs play fundamentally different roles in the considered system. UAV nodes are decision-making agents with mobility, battery, computation, and inter-UAV coordination states, whereas MT nodes represent service demand, task size, location, and A2G channel conditions. A homogeneous graph encoder would force these different entities and relations to share the same feature transformation and attention semantics, thereby mixing UAV-UAV cooperation with UAV-MT service-demand aggregation. In contrast, the HAN encoder uses type-specific transformations and meta-path-based aggregation to separately capture UAV-UAV and UAV-MT interactions. In the feature transformation stage, considering that UAV and MT nodes have different feature dimensions and physical meanings, we employ type-specific linear transformation matrices to map them into a common feature space. This step eliminates feature discrepancies among heterogeneous nodes. The node feature transformation is expressed as
\begin{equation}
	\mathbf{h^{\prime}}_i = \mathbf{W}_{\phi} \mathbf{h}_i + \mathbf{b}_{\phi}, \quad \phi \in \{\mathrm{UAV}, \mathrm{MT}\},
\end{equation}
where $\mathbf{h}_i$ denotes the original feature vector of node $i$, $\mathbf{W}_{\phi}$ and $\mathbf{b}_{\phi}$ represent the weight matrix and bias vector for node type $\phi$, and $\mathbf{h'}_i$ is the transformed feature.

For a given meta-path $\Phi$, we need to distinguish the importance of different neighbor nodes to the central node. To this end, we introduce a self-attention mechanism to compute attention coefficients $e_{ij}^{\Phi}$ based on concatenated features, that is,
\begin{equation}
	e_{ij}^{\Phi} = \mathrm{LeakyReLU}(\mathbf{a}_{\Phi}^T [\mathbf{h^{\prime}}_i \| \mathbf{h^{\prime}}_j]),
\end{equation}
where $\mathbf{a}_{\Phi}$ is the attention vector for meta-path $\Phi$, and $\|$ denotes feature concatenation. To stabilize the learning process, we extend this to multi-head attention, computing $K$ independent attention mechanisms and concatenating their results. The attention coefficients are normalized as
\begin{equation}
	\alpha_{ij}^{\Phi} = \frac{\exp(e_{ij}^{\Phi})}{\sum_{k \in \mathcal{N}_i^{\Phi}} \exp(e_{ik}^{\Phi})},
\end{equation}
where $\mathcal{N}_i^{\Phi}$ represents the set of neighbor nodes of node $i$ under meta-path $\Phi$. Based on the normalized attention weights, we aggregate neighbor features as
\begin{equation}
	\mathbf{z}_i^{\Phi} = \sigma\left(\sum_{j \in \mathcal{N}_i^{\Phi}} \alpha_{ij}^{\Phi} \mathbf{h^{\prime}}_j\right),
\end{equation}
where $\sigma(\cdot)$ is an activation function, and $\mathbf{z}_i^{\Phi}$ denotes the aggregated feature of node $i$ under meta-path $\Phi$.

Different meta-paths capture distinct semantic information of the network, which needs to be fused to form the final node representation. We integrate an attention module at the semantic level to dynamically capture the corresponding weights of various meta-paths, which can be modeled as
\begin{equation}
	w^{\Phi} = \frac{1}{|\mathcal{V}|} \sum_{i \in \mathcal{V}} \mathbf{q}^T\tanh(\mathbf{W}_s \mathbf{z}_i^{\Phi} + \mathbf{b}_s),
\end{equation}
where $\mathbf{q}$ is a semantic attention query vector, and $\mathbf{W}_s$ and $\mathbf{b}_s$ are learnable parameters. The weights for all meta-paths are normalized as
\begin{equation}
	\beta^{\Phi} = \frac{\exp(w^{\Phi})}{\sum_{\Phi' \in \Psi} \exp(w^{\Phi^{\prime}})},
\end{equation}
where $\Psi$ denotes the set of all meta-paths. Then, the fused feature representation of node $i$ is
\begin{equation}
	\mathbf{z}_i(t) = \sum_{\Phi \in \Psi} \beta^{\Phi} \mathbf{z}_i^{\Phi}.
\end{equation}
Finally, we can concatenate the feature vectors of all nodes into a global vector $\mathbf{Z}(t)$ in time slot $t$.

By leveraging the aforementioned hierarchical attention mechanism, HAN possesses the capability to adaptively discern the intrinsic relationships between UAVs and various network entities. This enables agents to prioritize more critical entities, thereby significantly improving the effectiveness of their decision-making policies. The learned node-level attention weights also provide an interpretable view of the decision process. In particular, the UAV-to-UAV attention weights indicate how much importance each UAV agent assigns to its neighboring UAVs when constructing the topology-aware state embedding. Therefore, visualizing these weights can help explain whether HAN-MATD3 treats all neighboring UAVs uniformly or learns non-uniform cooperative relationships under dynamic topology.

\subsection{HAN-MATD3}

We propose the HAN-MATD3 algorithm as Fig.~\ref{fig:algorithm}, which integrates the HAN encoder into the MATD3 framework under the CTDE paradigm. In our framework, each UAV agent $i$ maintains one actor network $\pi_i(\cdot|\theta_i^{\mu})$ and two critic networks $Q_i^{(1)}(\cdot|\theta_i^{Q_1})$ and $Q_i^{(2)}(\cdot|\theta_i^{Q_2})$. With the two critic networks, each UAV can deal with the overestimation problem of the Q-values in the one-critic framework. Furthermore, we incorporate the target actor $\pi_i^{\prime}(\cdot|\theta_i^{\mu^{\prime}})$ alongside target critics $\{Q_i^{(j)}(\cdot|\theta_i^{Q_j})\}_{j=1,2}$ to ensure a more stable training process.

The objective of HAN-MATD3 is to optimize the policy to maximize the expected cumulative reward, which is formulated as
\begin{equation}
	\max_\pi\mathbb{E} \left[ \sum_{t=0}^{T} \gamma^t r(t) \right],
\end{equation}
where $\mathbb{E}$ represents the expectation under the policy $\pi$. Let $G(t)$ denote the heterogeneous graph constructed in time slot $t$. After HAN encoding, each UAV agent $i$ obtains an embedding vector $\mathbf{z}_i(t)$, and we denote all agents' embeddings as $\mathbf{Z}(t)$ and the joint action as $\mathbf{a}(t)$. Under CTDE, the decentralized actor of agent $i$ only takes $\mathbf{z}_i(t)$ (derived from its local observation graph) as input, while the centralized critic can access the global embedding $\mathbf{Z}(t)$ and joint action $\mathbf{a}(t)$ during training.

\subsubsection{Actor}
The actor network of agent $i$ operates in a decentralized manner, taking only its local embedding $\mathbf{z}_i(t)$ as input to output the deterministic action as
\begin{equation}
	a_i(t)=\pi_i(\mathbf{z}_i(t);\theta_i^{\mu}).
\end{equation}
During training, exploration noise $\eta_i(t)$ is added to the action. To reduce variance and prevent the exploitation of Q-function errors, the target action for the next state is smoothed by adding clipped noise as
\begin{equation}
	a_i^{\prime}(t+1)=\pi_i^{\prime}(\mathbf{z}_i(t+1);\theta_i^{\mu'})+\epsilon_i,
\end{equation}
where $\epsilon_i$ represents the exploratory noise drawn from a Gaussian distribution. In UAV trajectory control, exploration is difficult because the policy must adapt to dynamic interference, time-varying A2G channel quality, service handover, and a hybrid high-dimensional action space. Cellular-connected UAV path planning studies, such as \cite{Qi2022QiERUAV}, show that improved replay and exploration mechanisms can enhance trajectory learning stability in dynamic wireless environments. In this work, HAN-MATD3 improves learning from both representation and policy-update perspectives: HAN converts the dynamic heterogeneous topology into compact embeddings, while MATD3 uses clipped double-Q learning, delayed policy updates, and target policy smoothing to reduce Q-value overestimation and unstable policy updates. The actor parameters are updated by maximizing the Q-value estimated by the first critic as
\begin{equation}
	J(\theta_i^{\mu}) = \mathbb{E}\left[Q_i^{(1)}(\mathbf{Z}(t),\mathbf{a}(t))|_{a_i=\pi_i(\mathbf{z}_i)}\right].
\end{equation}
Furthermore, the actor and target networks are updated with a latency interval $d$ to stabilize the learning process.

\subsubsection{Critic}
The critic networks operate in a centralized manner during training, taking the global state information $\mathbf{Z}(t)$ and joint action $\mathbf{a}(t)$ as inputs. A replay buffer $\mathcal{D}$ stores transitions $(\mathbf{Z}(t),\mathbf{a}(t),r(t),\mathbf{Z}(t+1))$. The target Q-value is calculated via the clipped double-Q learning mechanism, which is expressed as
\begin{equation}
	y_i(t) = r(t)+\gamma \min_{j=1,2} Q_i^{\prime(j)}(\mathbf{Z}(t+1),\mathbf{a}'(t+1)).
\end{equation}

The parameters of both critics are updated by minimizing the mean squared Bellman error as
\begin{equation}
	\mathcal{L}(\theta_i^{Q_j}) = \mathbb{E}_{\mathcal{D}}\left[(Q_i^{(j)}(\mathbf{Z}(t),\mathbf{a}(t)) - y_i(t))^2\right].
\end{equation}
Finally, the target networks are softly updated using a coefficient $\tau$ after every actor update. The detailed procedure is outlined in Algorithm~\ref{alg:HAN-MATD3}.
\begin{algorithm}
	\caption{HAN-MATD3 for Multi-UAV Semantic MEC}
	\label{alg:HAN-MATD3}
	\begin{algorithmic}[t]
		\State \textbf{Input:} System parameters and hyperparameters.
		\State \textbf{Output:} Trained actor policies for all UAVs.
		\State Initialize actor and critic networks, target networks, and replay buffer.
		\State Initialize HAN encoder parameters.
		\For{\textbf{each} episode}
		\State Reset environment and initialize state.
		\For{time slot $t=1,\dots,T$}
		\State Construct heterogeneous graph $G(t)$ based on current state.
		\State Generate state embeddings $\mathbf{Z}(t)$ using HAN encoder.
		\State Select actions $\mathbf{a}(t)$ via actor policies with exploration noise.
		\State Execute joint action, observe reward and next state.
		\State Store transition tuple in replay buffer $\mathcal{D}$.
		\If{replay buffer size $\ge$ batch size}
		\State Sample a random mini-batch from $\mathcal{D}$.
		\State Compute target actions with policy smoothing noise.
		\State Update critic networks by minimizing Bellman error.
		\If{delayed update interval reached}
		\State Update actor policies via deterministic policy gradient.
		\State Soft update target networks for actors and critics.
		\EndIf
		\EndIf
		\EndFor
		\EndFor
	\end{algorithmic}
\end{algorithm}
\subsection{Complexity Analysis}

Our presented HAN-MATD3 framework incurs computational overhead that primarily stems from the HAN-based heterogeneous graph encoder and the actor-critic updates. The HAN encoder's complexity per time slot is $\mathcal{O}\left(L_{\text{HAN}}\left(|\mathcal{V}| d_{\text{in}} d_z + H d_z \sum_{\Phi\in\Psi} |\mathcal{E}_{\Phi}| + |\Psi||\mathcal{V}| d_z\right)\right)$, where $L_{\text{HAN}}$ denotes the number of HAN layers, $d_{\text{in}}$ and $d_z$ represent the dimensions of the input node features and the output node embeddings, respectively. Furthermore, $H$ indicates the number of attention heads, and $|\mathcal{E}_{\Phi}|$ stands for the number of edges associated with a specific meta-path $\Phi$. Then, $|\mathcal{V}|$ and $|\mathcal{E}_{\Phi}|$ denote the number of nodes and meta-path-based edges, respectively, indicating that the encoding cost scales linearly with the graph size and connectivity. For the model training, the dominant complexity per iteration is $\mathcal{O}\left(2 N B_k C_{Q} + \frac{N B_k C_{\mu}}{d}\right)$, driven by the mini-batch updates of $N$ agents' critics and the delayed updates of their actors, where $B_k$ is the batch size, $C_{Q}$ and $C_{\mu}$ represent the forward/backward costs of the networks. 

The centralized critic becomes more expensive when the number of UAV agents increases, because the critic input includes the global graph embedding and the joint action. Therefore, the current framework is more suitable for small- and medium-scale cooperative UAV-MEC scenarios. In large-scale deployments, the communication and computation overhead can be reduced by constructing local neighborhood graphs, sampling meta-path-based neighbors, sharing actor parameters among homogeneous UAVs, or replacing the centralized critic with clustered or distributed critics, while preserving the topology-aware representation ability of HAN.

In practical semantic communication systems, semantic similarity depends on the semantic encoder/decoder architecture, task modality, training dataset, and time-varying wireless channels. The learning-based UAV radio resource management studies, such as \cite{Li2023RRMUAV}, also show that dynamic channel conditions affect the stability of communication resource allocation policies. Therefore, the ABG-based formulation in this paper is treated as a tractable system-level abstraction for joint UAV-MEC optimization. The task-oriented semantic encoders and online semantic-quality estimation remain important extensions of the proposed framework.

\section{Numerical Results}
\begin{table}[t]
	\centering
	\caption{Key environment parameters.}
	\label{tab:result}
	\begin{tabularx}{\linewidth}{Xl} 
		\toprule
		Parameters & Values (Unit) \\
		\midrule
		Number of UAVs & 3 \\
		Number of MTs & 10 \\
		Duration of service period & 200 (s) \\
		Duration of each time slot $\Delta t$ & 1 (s)\\
		Maximum altitude of flight $z_{\max}$ & 120 (m) \\
		Minimum altitude of flight $z_{\min}$ & 60 (m) \\
		Maximum velocity $v_{\max}$ & 50 (m/s) \\
		Minimum safety distance $d_{\min}$ & 10 (m) \\
		Carrier frequency $f_c$ & 2 (GHz) \\
		Noise power $\sigma^2$ & -174 (dBm) \\
		Transmit power of MT $p_m$ & 0.1 (W) \\
		Transmit power of UAV $p_n$ & 0.5 (W) \\
		CPU frequency of MT $f_m$ & 1 (GHz) \\
		CPU frequency of UAV $f_n$ & 5 (GHz) \\
		Effective switched capacitance coefficient $\kappa$ & $10^{-28}$\\
		Computation cycles per bit $C^{\mathrm{com}}$ & 1000 (cycles/bit) \\
		Encoding cycles per bit $C^{\mathrm{enc}}$ & 500 (cycles/bit) \\
		Decoding cycles per bit $C^{\mathrm{dec}}$ & 200 (cycles/bit) \\
		Mean task size $u_m$ & 4 (Mbits) \\
		Semantic threshold $\delta_{\text{th}}$ & 0.80 \\
		Semantic settings $g_1(n_{b}),g_2$ & 0.92, 2.08 \\
		Semantic settings $g_3,g_4$ & 7.28, 0.97 \\
		UAV settings $P_0,P_1$ & 79.86, 88.63 (Watt) \\
		UAV settings $U_\mathrm{tip},v_{0}$ & 120, 4.03 (m/s)\\
		UAV settings $\rho$ & 1.225 (kg/m$^3$)\\
		UAV settings $A$ & 0.503 (m$^2$)\\
		UAV settings $s$ & 0.05\\
		Weighting factors $\omega_L,\omega_E,\omega_\delta$ & 1/3, 1/3, 1/3 \\
		Battery capacity $E^{\mathrm{batt}}$ & $5\times 10^5$ (J) \\
		Penalty coefficients $\vartheta_1,\vartheta_2,\vartheta_3,\vartheta_4$ & 0.5, 0.5, 1.0, 2.0 \\
		LoS probability parameters $\kappa_0,\kappa_1$ & 9.61, 0.16 \\
		Path loss exponent $\lambda$ & 2.3 \\
		Excessive path loss coefficients $\eta_{\mathrm{LoS}},\eta_{\mathrm{NLoS}}$ & 1, 20 \\
		UAV-UAV communication range $R_{\mathrm{U2U}}$ & 500 (m) \\
		UAV-MT coverage radius $R_{\mathrm{U2M}}$ & 500 (m) \\
		\bottomrule
	\end{tabularx}
\end{table}
In this section, we evaluate the performance of the proposed HAN-MATD3 algorithm through simulation experiments. We first introduce the simulation setup, and then compare our method with several baseline algorithms.

\subsection{Experimental Setup}

The simulation framework is implemented using Python 3.13 and PyTorch 2.10.0. In our setup, the initial locations of both MTs and UAVs are randomly generated within a 1000 m × 1000 m terrestrial region. The operational altitude for UAVs is restricted to the range of [60, 120] m. Furthermore, the mobility of MTs follows the Gauss-Markov model. A comprehensive summary of other key environmental parameters is provided in Table~\ref{tab:result}.

\subsection{Baselines}

To validate the effectiveness of the proposed algorithm, we compare HAN-MATD3 with the following baseline algorithms:

MADDPG~\cite{Multi2021Wang}: Multi-agent deep deterministic policy gradient, an off-policy actor-critic framework using fully connected neural networks to process observations, is trained under the CTDE paradigm but without considering graph structural information.

MASAC~\cite{Kim2024MASAC}: A maximum entropy-based multi-agent continuous control algorithm. It utilizes entropy regularization to enhance exploration and relies on fully connected networks for observation processing.

MATD3~\cite{Zhao2022MATD3}: Multi-agent twin delayed DDPG is an algorithm that uses fully connected neural networks to process observations and inherits TD3's stability improvements, but does not leverage graph structures to model inter-agent relationships.

NSGA-II~\cite{Zhu2022NSGA}:  A popular evolutionary algorithm that optimizes the decision variables via heuristic search. It is executed in a slot-by-slot manner to handle the dynamic environment.

All DRL algorithms adopt the same hyperparameter settings to ensure fair comparison: learning rates $\mu_a = 10^{-4}$ (Actor) and $\mu_c = 10^{-3}$ (Critic), discount factor $\gamma = 0.99$, soft update coefficient $\tau = 0.005$, batch size $B_k$ = 256, replay buffer size $|\mathcal{D}| = 10^6$, delayed update interval $d = 2$. The parameters in the ABG empirical model are derived from \cite{Ma2025semantic}.

To further quantify the contribution of key internal modules in HAN-MATD3, we conduct an ablation study on heterogeneous modeling and semantic-level fusion under the same experimental settings. The considered ablation variants are as follows:

w/o Heterogeneity: UAVs and MTs are treated as homogeneous nodes. This variant removes type-specific transformations and heterogeneous interaction modeling, and is used to evaluate the contribution of heterogeneous graph modeling.

w/o Semantic Fusion: The UAV-UAV and UAV-MT meta-path attention outputs are fused by fixed averaging instead of learnable semantic-level fusion. This variant evaluates the contribution of adaptive meta-path weighting.

All ablation variants adopt the same environment settings, reward function, action space, and training hyperparameters as the full HAN-MATD3 algorithm to ensure fair comparison.

\subsection{Overall Results}
\begin{figure}[t]
	\centering
	\includegraphics[width=\columnwidth]{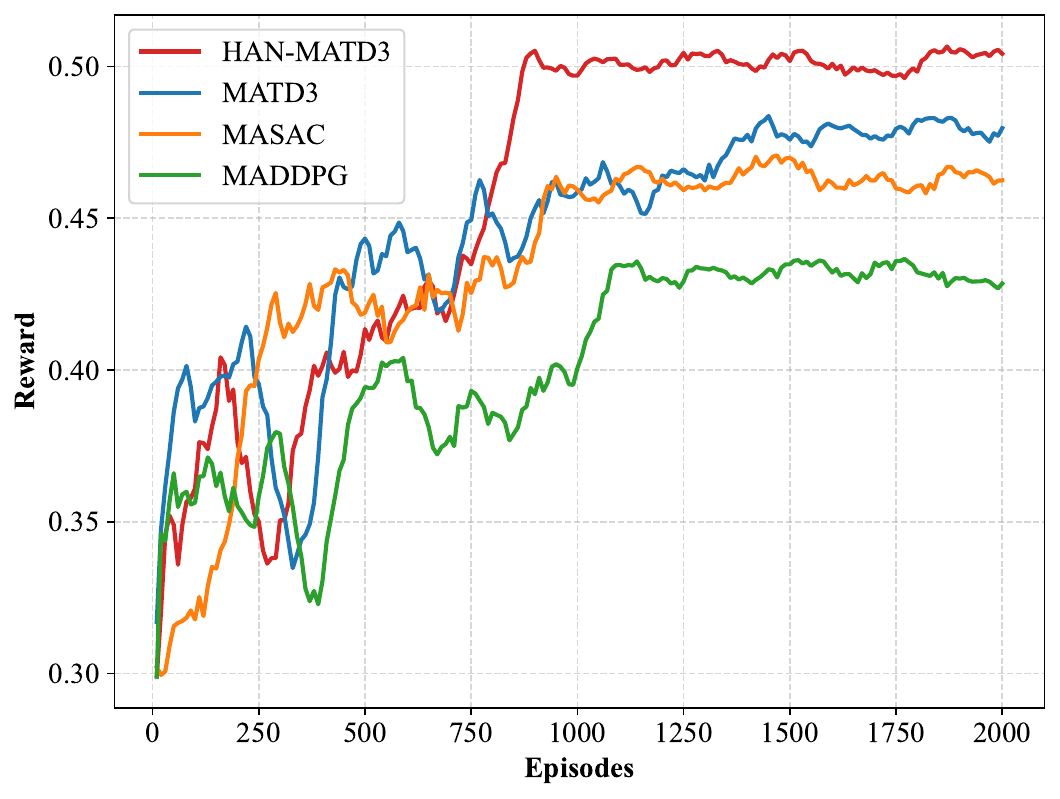}
	\caption{Convergence of reward.}
	\label{fig:reward}
\end{figure}

\begin{figure}[t]
	\centering
	\includegraphics[width=\columnwidth]{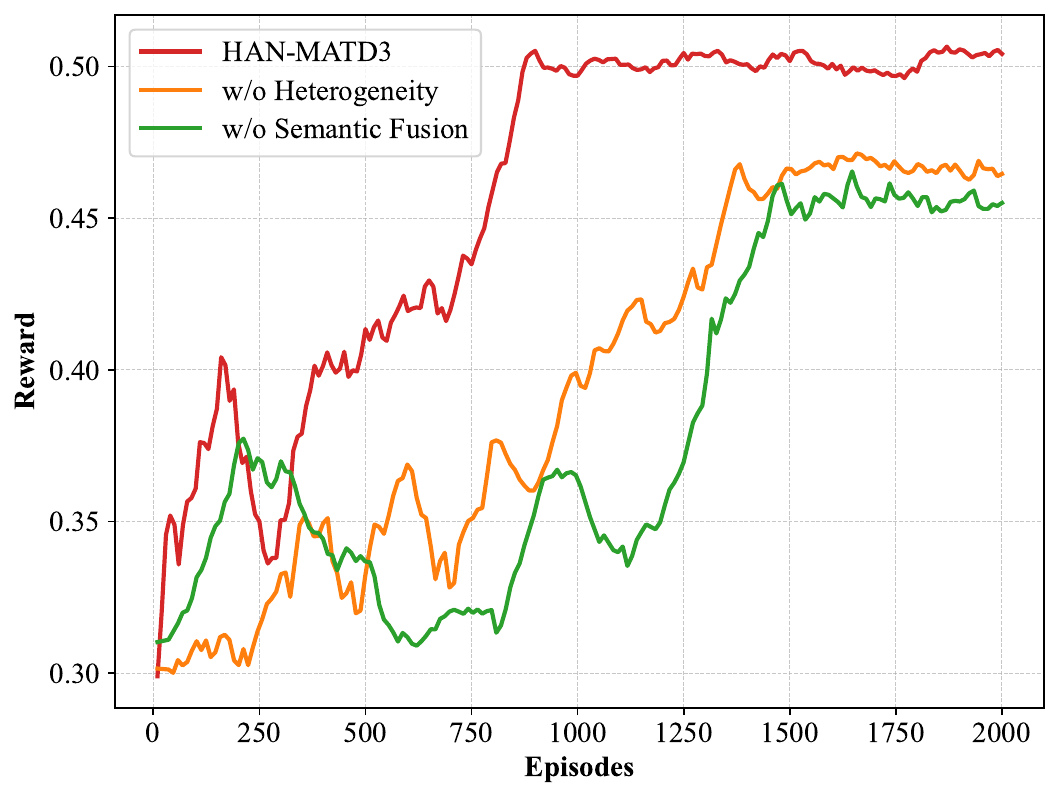}
	\caption{Ablation study on the convergence performance of HAN-MATD3.}
	\label{fig:ablation_reward}
\end{figure}

\subsubsection{Convergence analysis}
Fig.~\ref{fig:reward} illustrates the convergence curves of the total reward for HAN-MATD3 and three baseline algorithms throughout the training process. It can be observed that as the number of episodes increases, the reward values of all algorithms exhibit an upward trend and gradually converge to a stable state. Among all the compared algorithms, HAN-MATD3 exhibits the superior convergence performance and stability. Specifically, the convergence curve of HAN-MATD3 consistently remains above those of other baselines, finally stabilizing at approximately 0.505. Compared to the second-best MATD3 algorithm, HAN-MATD3 achieves a performance gain of about 5.3\%. Furthermore, compared to the traditional MADDPG algorithm, the improvement is as high as 17.7\%.
The rapid convergence of the proposed algorithm is attributed to the incorporation of the HAN. The HAN can accurately capture the semantic dependencies among heterogeneous nodes in the edge environment, thereby providing a more robust state representation for the reinforcement learning agents. Simultaneously, the integration with the MATD3 mechanism effectively mitigates the Q-value overestimation issue, further ensuring the stability of policy updates.

Fig.~\ref{fig:ablation_reward} shows that the full HAN-MATD3 achieves the highest final reward and converges to a stable reward level, which verifies the benefit of jointly using topology-aware graph representation, heterogeneous modeling, and semantic-level fusion. In the early training stage, the ablated variants exhibit larger fluctuations because removing either type-specific modeling or learnable semantic-level fusion weakens the state representation used by the agents during policy exploration. After the policies become stable, the fluctuations of all curves decrease. The w/o Heterogeneity variant reaches a lower reward than the full model because UAVs and MTs are not explicitly distinguished, although it still retains graph aggregation, semantic fusion, and the MATD3 learning framework. The w/o Semantic Fusion variant converges more slowly and reaches a slightly lower final reward because fixed average fusion cannot adaptively weight different meta-path/relation semantics under dynamic network topology and channel conditions. These results confirm that heterogeneous modeling and semantic-level fusion both contribute to the convergence performance of the proposed framework.

\begin{figure}[t]
	\centering
	\includegraphics[width=\columnwidth]{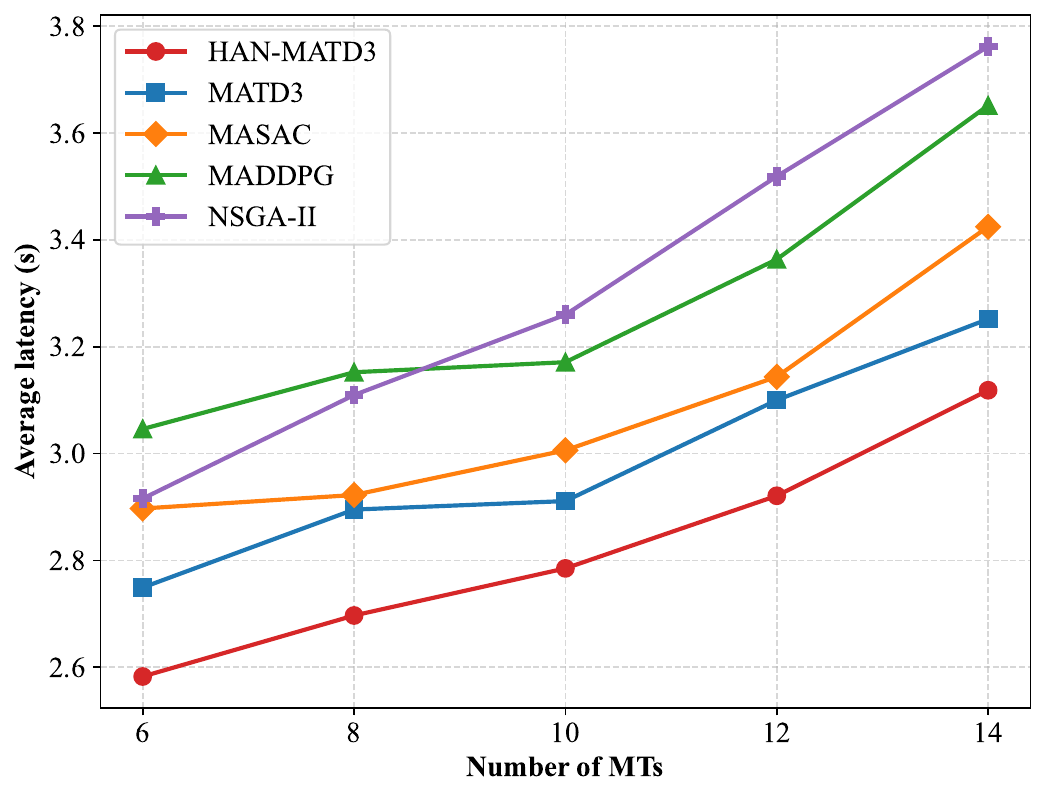}
	\caption{The effect of the number of MTs on average latency under a fixed number of UAVs.}
	\label{fig:mt_lat}
\end{figure}

\begin{figure}[t]
	\centering
	\includegraphics[width=\columnwidth]{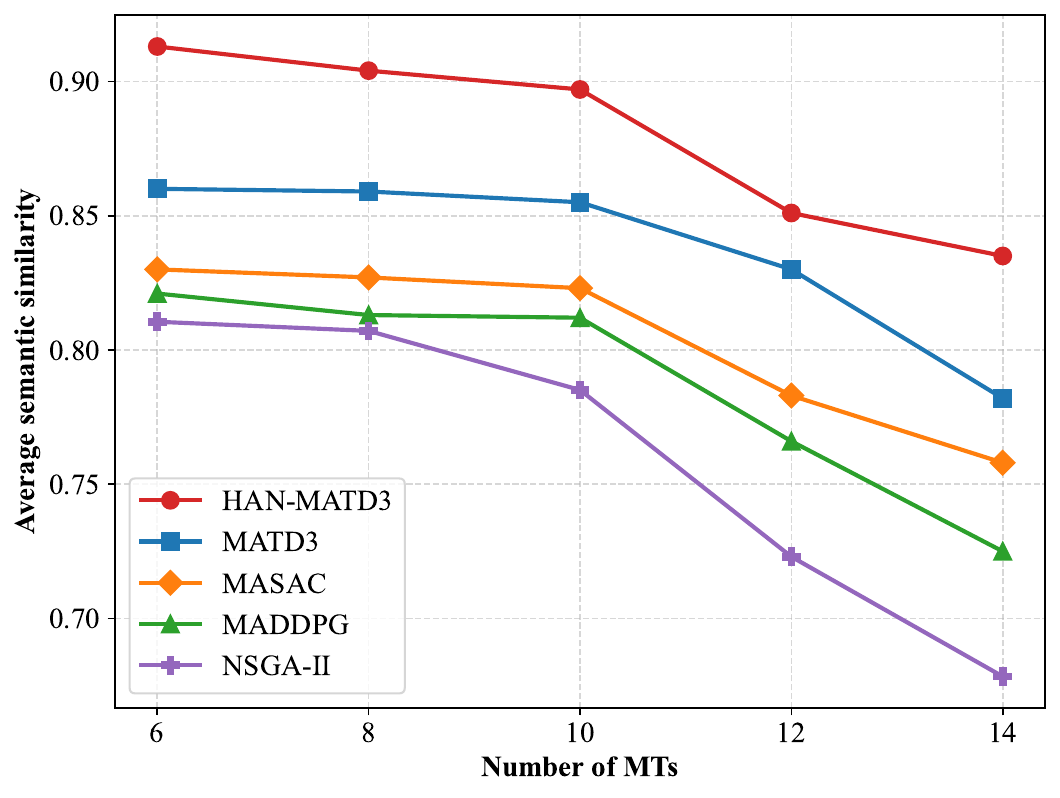}
	\caption{The effect of the number of MTs on average semantic similarity under a fixed number of UAVs.}
	\label{fig:mt_sem}
\end{figure}

\begin{figure}[t]
	\centering
	\includegraphics[width=\columnwidth]{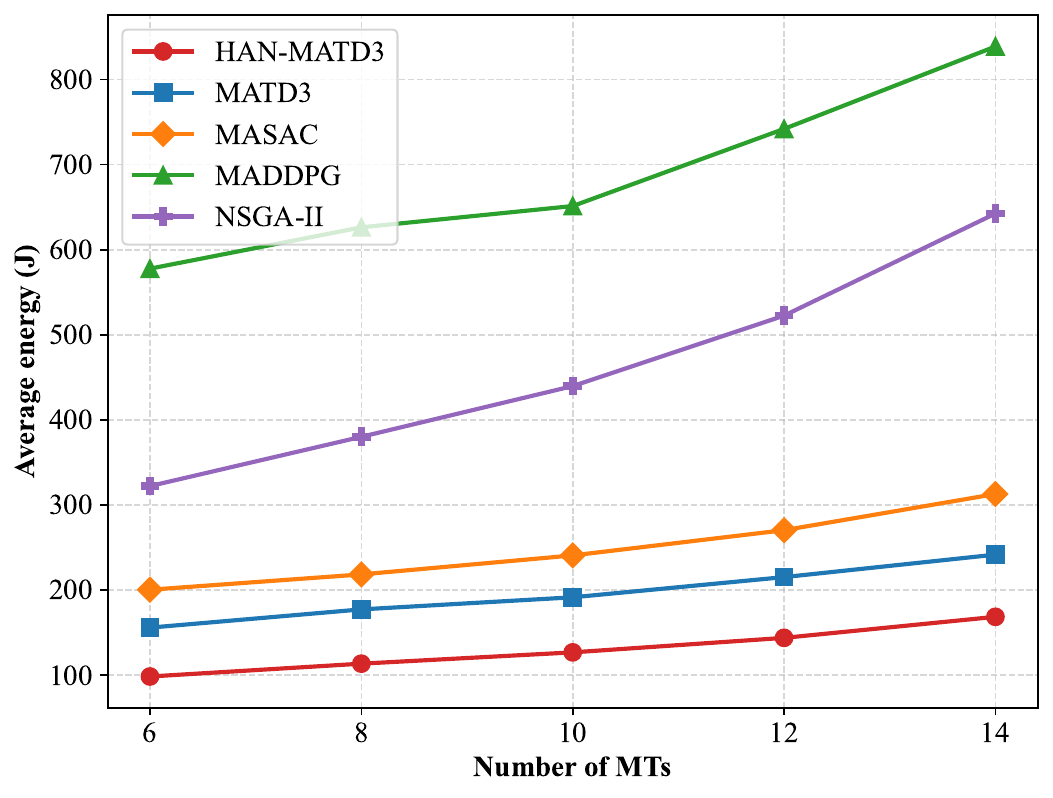}
	\caption{The effect of the number of MTs on average energy per step under a fixed number of UAVs.}
	\label{fig:mt_eng}
\end{figure}

\begin{figure}[t]
	\centering
	\includegraphics[width=\columnwidth]{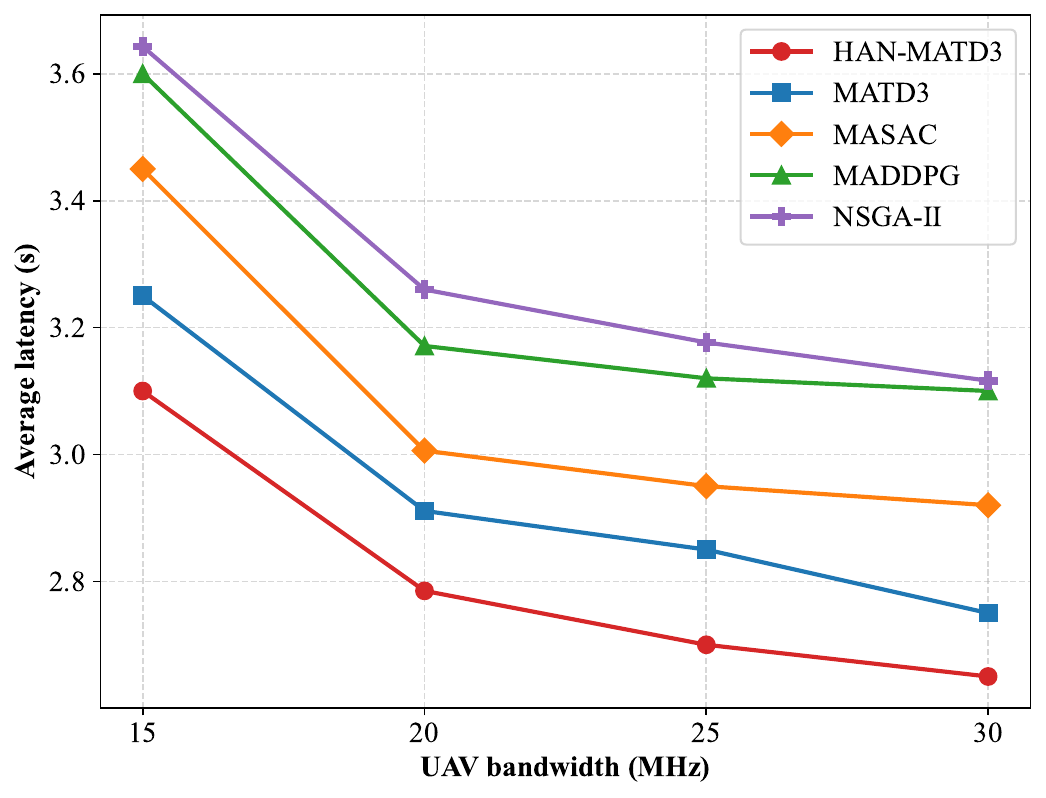}
	\caption{The effect of different bandwidths on average latency.}
	\label{fig:b_lat}
\end{figure}

\begin{figure}[t]
	\centering
	\includegraphics[width=\columnwidth]{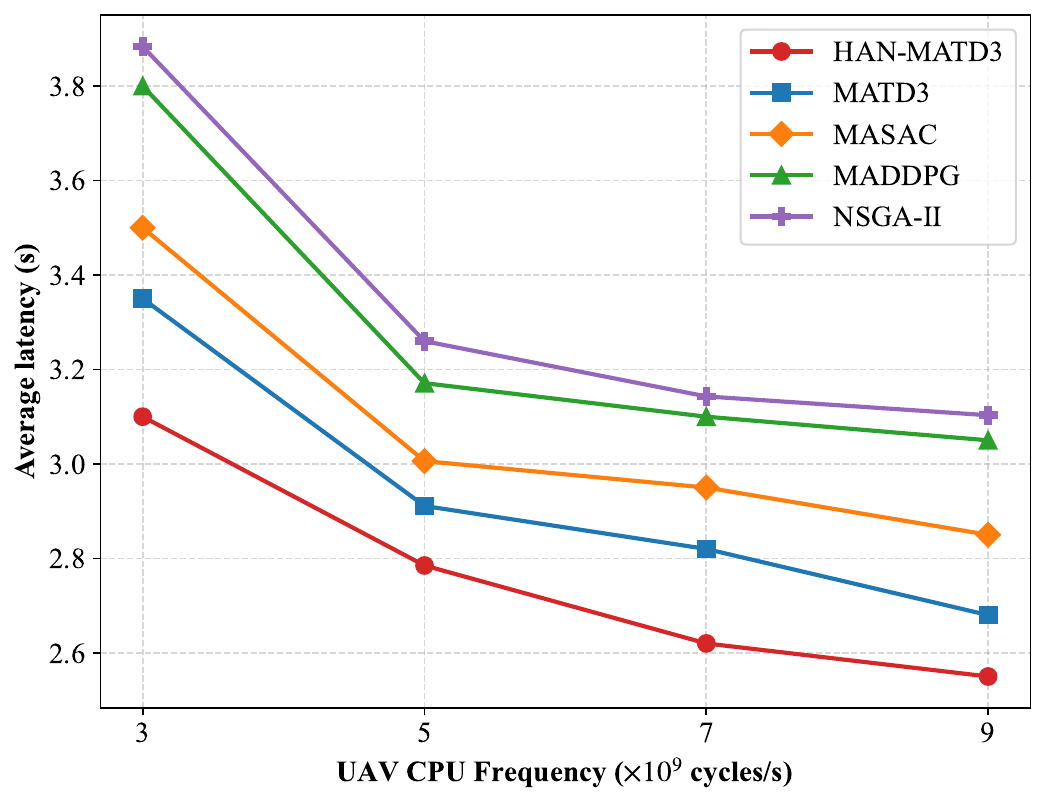}
	\caption{The effect of different UAV CPU computation frequencies on average latency.}
	\label{fig:uav_f_lat}
\end{figure}

\subsubsection{Performance evaluation}
Fig.~\ref{fig:mt_lat} illustrates the impact of the number of MTs on the average latency. As the number of MTs increases from 6 to 14, the average latency for all algorithms exhibits an upward trend. This performance gap can be attributed to the intensified competition for limited spectrum and computation resources as MTs density grows, which subsequently prolongs the queuing time for both processing and transmission. Among all evaluated algorithms, HAN-MATD3 consistently maintains the lowest latency. Specifically, when the number of MTs reaches 14, HAN-MATD3 achieves a latency of 3.1185 s. This performance surpasses that of MATD3, MASAC, and MADDPG, which record latencies of 3.2518 s, 3.4246 s, and 3.6516 s, respectively. Furthermore, HAN-MATD3 demonstrates a significant advantage over the traditional NSGA-II algorithm, which yields the highest latency of 3.7620 s under high traffic loads. These results validate that by jointly optimizing resource allocation and trajectory control, HAN-MATD3 effectively mitigates interference and congestion in multi-MT scenarios, thereby facilitating lower data transmission latency.

As illustrated in Fig.~\ref{fig:mt_sem}, as the number of MTs increases, the semantic similarity of all algorithms exhibits a downward trend. This degradation is primarily attributed to the reduced SINR caused by multi-MT interference, which subsequently increases the error rate of semantic decoding. Nevertheless, the HAN-MATD3 algorithm consistently demonstrates the optimal semantic similarity across various MT densities. Specifically, when the number of MTs reaches 14, HAN-MATD3 maintains a semantic similarity of 0.835. In contrast, the values for MATD3, MASAC, and MADDPG decline to 0.782, 0.758, and 0.725, respectively. Notably, the NSGA-II algorithm exhibits the most significant performance deterioration, yielding a score of only 0.678 under the same conditions. These results validate that HAN-MATD3 is capable of more effectively extracting and transmitting key semantic features, ensuring high-quality semantic information interaction even in high-load network environments.

Fig.~\ref{fig:mt_eng} compares the average energy consumption under different MT densities. As the number of MTs increases, all algorithms consume more energy because more tasks need to be processed and transmitted. HAN-MATD3 consistently achieves the lowest energy consumption, reaching 168.30 J when the number of MTs is 14. Although NSGA-II consumes less energy than MADDPG under this setting, its lower energy cost is accompanied by higher latency and lower semantic similarity, indicating an unbalanced latency-energy-semantic tradeoff. In contrast, HAN-MATD3 achieves lower energy consumption while maintaining better service quality.

Fig.~\ref{fig:b_lat} and Fig.~\ref{fig:uav_f_lat} illustrate the impact of A2G uplink bandwidth and UAV CPU computation frequency on average latency, respectively. As bandwidth increases, transmission latency decreases, while higher CPU frequency accelerates task execution. Across all tested configurations, HAN-MATD3 achieves the lowest latency by using heterogeneous graph attention for topology-aware resource allocation. Under the baseline setting with 20 MHz bandwidth and 5 GHz CPU frequency, HAN-MATD3 achieves a latency of 2.785 s, corresponding to reductions of approximately 12.2\% over MADDPG and 14.6\% over NSGA-II. These results show that HAN-MATD3 can better alleviate both communication and computation bottlenecks.

\subsubsection{Attention visualization}
\begin{figure}[t]
	\centering
	\includegraphics[width=0.45\textwidth]{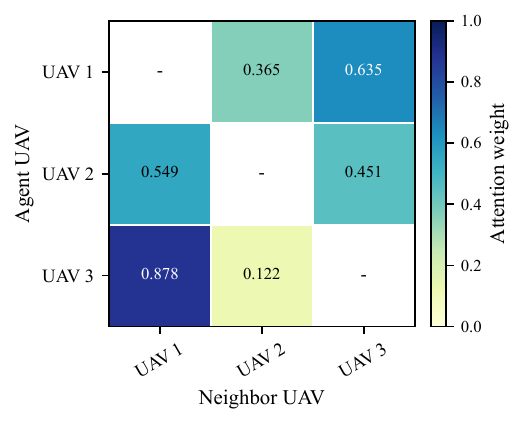}
	\caption{Visualization of learned UAV-to-UAV node-level attention weights.}
	\label{fig:attention_uav}
\end{figure}

To further interpret the topology-aware representation learned by HAN, we visualize the UAV-to-UAV node-level attention weights of the trained encoder. The averaged attention weights are shown in Fig.~\ref{fig:attention_uav}. Each row represents the attention distribution assigned by one UAV to its neighboring UAVs, and the diagonal entries are omitted because self-attention is not considered in this visualization. The non-diagonal entries show that different UAVs assign non-uniform importance to their neighbors. For example, UAV 1 assigns the weights of 0.365 and 0.635 to UAV 2 and UAV 3, respectively, while UAV 3 assigns a dominant weight of 0.878 to UAV 1. These results suggest that the HAN encoder learns topology-dependent cooperative relationships instead of simply averaging neighboring UAV information.

\section{Conclusion}

In this paper, we have investigated the joint optimization of computation offloading, task migration, resource allocation, and UAV trajectory control in semantic communication-enabled multi-UAV MEC systems. Specifically, we have formulated a joint optimization problem aiming to maximize semantic similarity while minimizing task execution latency and system energy consumption. To address the challenges posed by dynamic network topology and complex multi-agent coordination, we have proposed a novel HAN-MATD3 framework. By integrating the HAN with the MATD3 algorithm, the proposed method effectively captures the intricate dependencies among heterogeneous nodes and learns robust cooperative policies. Extensive simulation results have demonstrated that our scheme outperforms existing baselines. Notably, it has achieved a significant reduction in service latency and energy consumption, along with an improvement in semantic similarity, validating its superiority in resource-constrained edge environments. Subsequent research will adapt our framework to reconfigurable intelligent surface-assisted scenarios to further enhance link quality. Additionally, addressing the limitation of centralized training costs, we will explore fully decentralized graph learning approaches to further reduce communication overhead and improve scalability.

\bibliographystyle{IEEEtran}
\bibliography{reference.bib}  

\end{document}